\documentclass[10pt,apsrev4-1,prl,reprint,twocolumn,amsmath,superscriptaddress,preprintnumbers,floatfix,nofootinbib,showpacs]{revtex4-1}
\usepackage{siunitx}
\usepackage{graphicx}
\usepackage[dvipsnames]{xcolor}
\usepackage[bookmarks]{hyperref} 
\usepackage{tikz}
\usepackage{booktabs}
\usepackage{mathtools}
\usepackage{braket}
\usepackage[utf8]{inputenc}
\usepackage[T1]{fontenc}
\usepackage{natbib}
\usepackage{amssymb,latexsym}
\usepackage{wasysym}
\usepackage{soul}

\newcommand{\matr}[1]{\mathbf{#1}}

\begin{document}

\title{New approach to search for parity-even and parity-odd time-reversal violation\newline  beyond the Standard Model in a storage ring} 
	
\author{N. N. Nikolaev}
\affiliation{L.D. Landau Institute for Theoretical Physics, 142432 Chernogolovka, Russia}
\affiliation{Moscow Institute for Physics and Technology, 141700 Dolgoprudny, Russia}
 
\author{F. Rathmann}
\affiliation{Institut f\"ur Kernphysik, Forschungszentrum J\"ulich, 52425 J\"ulich, Germany }

\author{A. J. Silenko}
\affiliation{Laboratory of Theoretical Physics, Joint Institute for Nuclear Research,  Dubna, 141980 Russia}
\affiliation{Institute of Modern Physics, Chinese Academy of Sciences, Lanzhou 730000, China}
\affiliation{Research Institute for Nuclear Problems, Belarusian State University, Minsk 220030, Belarus}

\author{Yu. Uzikov}
\affiliation{Laboratory of Nuclear Problems, Joint Institute for Nuclear Research,  Dubna, 141980 Russia}
\affiliation{Dubna State University, Dubna, 141980 Russia}
\affiliation{Department of Physics, M.V. Lomonosov State University, Moscow, 119991 Russia}

\preprint{Draft \today}

\begin{abstract}
Time-reversal breaking and parity-conserving millistrong interactions, suggested in 1965, still remain a viable mechanism of CP-violation beyond the Standard Model. One of its possible manifestations is the T-odd asymmetry in the transmission of tensor-polarized deuterons through a vector-polarized hydrogen gas target. 
Upon the rotation of the deuteron polarization from the vertical direction into the ring plane, the T-odd asymmetries, odd against the reversal of the proton  polarization in the target, will continuously oscillate with first or second harmonics of the spin precession frequency.  The  Fourier analysis of the oscillating T-odd asymmetries allows for an easy separation from background persistent in conventional experiments employing static
 vector and tensor polarizations. 
\end{abstract}
\maketitle


In 1965 Okun\,\cite{Okun:1965tu}, Prentki \& Veltman\,\cite{Prentki:1965tt}, and Lee \& Wolfenstein\,\cite{Lee:1965hi} suggested millistrong, flavour and parity (P) conserving, but time-reversal (T) violating (TVPC) interactions, as a source of CP violation in neutral kaon decays. The Standard Model (SM) predicts the experimentally accessible CP violation only in flavour changing processes. In beyond  SM  (BSM) millistrong CP-odd interactions (MSCPV), T-violation in flavour \textit{conserving} strong and nuclear reactions, $\gamma$-transitions and $\beta$-decays is possible at the level of $\sim 10^{-3}$. An intriguing open issue is whether MSCPV can resolve the puzzle of the anomalously large baryon asymmetry of the Universe (for a review, see\,\cite{Garbrecht:2018mrp}). 

In this paper, we propose a new approach to the search for T-violation in double-polarized proton-deuteron ($pd$) transmission experiments. It is based on the rotation of the vertical polarization of stored deuterons into the ring plane. The in-plane vector and tensor polarizations  precess with the first or second harmonic of the idle spin-precession frequency. Their interactions with an internal polarized hydrogen target give rise to a hierarchy of polarization asymmetries. In the conventional search for T-violation in double-polarized $pd$ interactions\,\cite{Valdau:2011zz,Lenisa:2019cgb},  the static T-odd tensor asymmetry receives a systematic contribution from the hard-to-control background from vector polarized deuterons in the tensor polarized cell target\,\cite{Eversheim:2017zxl,Temerbayev:2015foa}(for a discussion of unwanted spin components in cell targets, see\,\cite{Rathmann:1998zz}). It is crucial that this daunting task of separating the T-violating  signal from the systematic background, can be  met with  Fourier analyses, which readily distinguishes  different polarization asymmetries by their oscillation frequency and parity with respect to the reversal of the proton  target polarization. This feature gives storage ring experiments with oscillating  in-plane polarizations a clear advantage over experiments with static deuteron and proton polarizations.
 
As a  proof-of-principle, in 2014 the JEDI collaboration for the first time  made use of the precessing in-plane deuteron polarization for a high precision measurement of the spin-precession frequency\,\cite{Bagdasarian:2014ega,Eversmann:2015jnk}. Subsequently, a feedback system to stabilize the spin-precession frequency was developed\,\cite{PhysRevLett.119.014801}. The possibility of fast reversal and rotation of the proton target polarization has been demonstrated, \textit{e.g.}, at IUCF\,\cite{Rathmann:1998zz}. Putting the oscillating polarization of stored protons to use in the search for P-violation in hadronic interactions at the NICA collider has recently been proposed in\,\cite{2019arXiv191110701K}. 
  A possibility to accelerate particles with precessing polarization has already been discussed in 2002 
 in the little noticed  publication
\,\cite{Sitnik}.

Our focus here will be on search for the MSCPV interaction via the TVPC asymmetry in the total cross section of double-polarized proton-deuteron ($pd$) scattering. Alongside the TVPC asymmetry, one can study in the same experimental setup the T-odd and P-odd (TVPV) asymmetry, which recently received much theoretical scrutiny (for an extensive discussion, see\,\cite{deVries:2020iea}). We show that, besides these two T-odd asymmetries, oscillating vector and tensor polarizations provide a new access to a whole family of further single- and double-polarization observables.

In the SM, flavour-conserving TVPC effects are only possible to the second order in weak interactions, and are far below the  reach of present experimental observations.
 An example of the second order observable is the much dicussed TVPV electric dipole moment of the neutron , expected in the SM at the level of $d_n \sim \SI{e-32}{e.cm}$ \cite{Khriplovich:1997ga}. The dimensional estimate from MSCPV amounts to $d_n \sim  \SI{e-24}{e.cm}$\,\cite{Okun:1967cxa,Conti:1992xn,Haxton:1994bq,RamseyMusolf:1999nk}. The best recent experimental upper bound is $d_n = (0.0\pm 1.1_{\rm stat} \pm 0.2_{\rm sys}) \cdot \SI{e-26}{e.cm} $\,\cite{PhysRevLett.124.081803}. However, in view of the complexity of the interplay of long-distance and short-distance effects, the present bounds on $d_n$ do not preclude MSCPV as a sourve of CP-violation\,\cite{Kurylov:2000ub,El-Menoufi:2016cfo}.

 Direct constraints on TVPC effects in flavour-conserving processes include tests of the detailed-balance principle\,\cite{Blanke:1983zz}, studies of the $\beta$-decay of polarized neutrons\,\cite{Mumm:2011nd}, transmission of polarized neutrons through a spin-aligned $^{165}$Ho target\,\cite{PhysRevC.55.2684}, and charge-symmetry breaking in spin observables of $pn$ scattering\,\cite{Simonius:1997sd}. According to \cite{Kurylov:2000ub,El-Menoufi:2016cfo}, these bounds  do not contradict expectations from the MSCPV interaction, and Kurylov et al.\ strongly emphasize the importance of direct searches for TVPC effects in scattering experiments\,\cite{Kurylov:2000ub}.

In transmission experiments, the best upper bound on the TVPC asymmetry
 has been achieved in interactions of 5.9 MeV polarized neutrons with tensor polarized $^{165}$Ho crystals, yielding $A_5 < 2.2\cdot10^{-5}$\,
\cite{PhysRevC.55.2684}. 
 The subsequent proposal  to search for the TVPC asymmetry in $pd$- scattering anticipated interactions of vertically polarized protons, stored in the COSY ring,
passing through a tensor-polarized deuterium storage cell target\,\cite{Valdau:2011zz}. The related theoretical estimates were performed for the energy range 100 to \SI{1000}{MeV}\,\cite{Beyer:1993zw,Uzikov:2015aua,Uzikov:2016lsc}. On a statistical basis, the TVPC asymmetry of $\sim 10^{-6}$ is within the reach of $pd$ transmission experiments\,\cite{Valdau:2011zz}. 


The further presentation is organized as follows. We start with a brief review of spin observables in $pd$ transmission experiments. Then, we derive the evolution equation of vector and tensor polarizations of deuterons when exposed to a resonant RF spin rotator, and report on explicit solutions for the precessing vector and tensor polarizations for the relevant initial vertical polarization. Next, we discuss how Fourier analyses furnish the determination of TVPC and TVPV asymmetries, and comment on the access to further spin observables, beyond the T-violating ones, using precessing polarizations. 
 
The  spin-dependent total $pd$ cross section is written as
\begin{equation}
\begin{split}
\label{sigmatotal}
\sigma_\text{tot} = & \,\sigma_0 + \sigma_\text{TT} \left[ \left( {\bf P}^\text{d} \cdot {\bf P}^\text{p} \right) - \left( {\bf P}^\text{d}\cdot {\bf k} \right) \left( {\bf P}^\text{p}\cdot {\bf k} \right) \right]\\
 & + \sigma_\text{LL} \left( {\bf P}^{\rm d} \cdot {\bf k} \right) \left( {\bf P}^\text{p}\cdot {\bf k} \right)
+\sigma_\text{T} T_{mn}k_m k_n \\
& + \sigma_\text{PV}^\text{p} \left( {\bf P}^\text{p} \cdot {\bf k} \right) + \sigma_\text{PV}^\text{d} \left( {\bf P}^\text{d} \cdot {\bf k} \right) \\
& +\sigma_\text{PV}^\text{T} \left( {\bf P}^\text{p} \cdot {\bf k} \right) T_{mn}k_m k_n \\
& + \sigma_\text{TVPV} \left( {\bf k} \cdot \left[ {\bf P}^\text{d} \times {\bf P}^\text{p} \right] \right) \\
& + \sigma_\text{TVPC} k_m T_{mn} \epsilon_{nlr}P_l^\text{p} k_r\, .
\end{split}
\end{equation}
Here ${\bf P}^{\rm d}$ and ${\bf P}^{\rm p}$ are  the vector polarizations of deuteron and proton, $T_{mn}$ is the tensor polarization of the deuteron and ${\bf k}$ is the unit vector along the collision axis. We chose the latter for the $z$-axis, the $y$-axis is orthogonal to the ring plane, so that  $T_{mn}k_mk_n =T_{zz}$, and 
\begin{equation}
  k_mT_{mn}\epsilon_{nlr}P_l^{\rm p} k_r
   = T_{xz}P_y^{\rm p}-T_{yz}P_x^{\rm p}\, . \label{eq:TVPC} 
  \end{equation}
In Eq.\,(\ref{sigmatotal}), the cross sections $\sigma_0$, $\sigma_\text{TT}$, $\sigma_\text{LL}$, and $\sigma_\text{T}$ correspond to ordinary P-even and T-even interactions, $\sigma_\text{PV}^\text{p}$, $\sigma_\text{PV}^\text{d}$, and $\sigma_\text{PV}^\text{T}$ are signals of the P-violation, and $\sigma_\text{TVPV}$ denotes the T- and P-violating one. The last term,  $\sigma_\text{TVPC}$, is the null observable for the TVPC  interaction\,\cite{Baryshevsky83,Barabanov:1986sz,Conzett:1992dn}. 
    
The TVPC and TVPV experiment with polarized deuterons in a storage ring with installed RF spin rotator is envisaged as follows. One starts with the injection of deuterons with vertical spin $\vec{S}$ 
and tuning conditions to provide a long spin coherence time\,\cite{PhysRevLett.117.054801,Guidoboni:2017ayl}. Throughout this paper, $\vec{S}(t)$ will stand for the time-dependent deuteron spin operator,  its expectation value,  $\vec{P}^{\rm d}(t)= \langle \vec{S}(t) \rangle$, is the polarization vector. The initial vertical vector polarization $ \langle S(0) \rangle = P_y^\text{d}(0) \vec{e}_y$ entails also the tensor polarization $T_{yy}(0)= \langle Q_{yy}(0)\rangle$, where the spin-tensor operator is defined as 
 \begin{equation}
 Q_{mn}(t) = S_m (t)S_n(t) +S_m(t) S_m(t) - \frac{2}{3} \vec{S}^2\delta_{mn} \, .\label{eq:Tensor}
 \end{equation} 
  
 The electric quadrupole moment of the deuteron is of the order of the deuteron magnetic moment times the size of the deuteron. In a storage ring, its interaction with the gradient of the motional electric field\,\cite{Pomeransky:1999ej} is some 13 to 14 orders in magnitude smaller than the interaction of the magnetic moment with the magnetic field. Consequently, the deuteron spin dynamics in a storage ring is entirely driven by the evolution of its vector polarization. 
 
 The angular velocity $\vec{\Omega}$ of the idle spin precession in the laboratory frame is given by the Thomas-Bargmann-Michel-Telegdi (T-BMT) equation\,\cite{PhysRevLett.2.435}. It is convenient to follow rotations of the spin with respect to the particle momentum, \textit{i.e.},  $\vec{\Omega}_s = \vec{\Omega} - \vec{\Omega}_r$, subtracting the cyclotron angular velocity $\vec{\Omega}_r$. A convenient quantity is the spin tune $\nu_s = \Omega_s/\Omega_r$. The idle spin-precession angle per turn equals $\theta_s = 2\pi \nu_s$. As a spin rotator, we consider an RF solenoid with magnetic field along the $z$-axis, tangential to the beam orbit. It is operated at the spin-precession frequency $f_s = \Omega_s/(2\pi) = \nu_s f_r$  (modulo to integers of the cyclotron frequency $f_r$). 
 
 As a function of the turn number $n$, the spin evolves stroboscopically: (1) it idly precesses about the vertical axis during a full revolution, and (2), while passing the RF solenoid, it acquires a turn-dependent rotation about the $z$-axis by the angle $\psi(n)= \psi_\text{RF} \cos(\theta_s n)$,  $\psi_\text{RF} \ll 1$. The polarimeter, located right behind the spin rotator, stroboscopically analyses the spin orientation once per turn. The RF spin rotator gives rise to the evolution of the envelope of the precessing polarization. The transition from the stroboscopic evolution to the continuous time dependence with $n = f_r\, t$ is furnished by the Bogoliubov-Krylov-Mitropolsky (BKM) averaging\,\cite{Bogolyubov}.  
          
Following the treatment described in\,\cite{Saleev:2017ecu}, extended to the $\mathbf{SO(3)}$ formalism, one can derive the BKM-averaged spin evolution in factorized form,
\begin{equation}  
     \begin{split}     
    \vec{S}(n)&=\matr{R}_\text{evol}(n) \vec{S}(0)\,,\\ 	
    \matr{R}_\text{evol}(n)&= \matr{R}_{\rm idle}(n)\matr{R}_{\rm env}(n)\, , 
    \end{split}
    \end{equation}
where the idle precession and the spin-envelope evolution matrices equal, respectively, 
 \begin{equation}
 \matr{R}_{\rm idle}(n) = 
 \begin{pmatrix}
 \cos\theta_s n & 0 & \sin \theta_sn \\
 0 &  1 & 0 \\
 -\sin\theta_s n & 0 & \cos \theta_s n
 \end{pmatrix}\,  , 
 \end{equation}
 \begin{equation}
 \matr{R}_{\rm env}(n) = 
 \begin{pmatrix}
 \cos \epsilon n  & \sin \epsilon n & 0 \\
 -\sin\epsilon n & \cos\epsilon n & 0 \\
 0 & 0 & 1
 \end{pmatrix}\, . \label{eq:Matrices}
 \end{equation}
The latter describes the  rotation of the spin envelope in the $xy$-plane by a constant angle $\epsilon = \psi_\text{RF}/2\, $ per turn, {\it i.e.,} with the spin-resonance tune (strength) $\nu_\text{res} = \epsilon/(2\pi)$. Then, the final spin-evolution operator becomes
  \begin{equation}
 \begin{split}
 &
 \matr{R}_\text{evol}(n) =\\
 &
 \begin{pmatrix}
 \cos \theta_s n \cdot \cos \epsilon n  & \cos \theta_s n \cdot \sin \epsilon n & \sin \theta_s n \\
 -\sin\epsilon n & \cos\epsilon n & 0 \\
 -\sin \theta_s n \cdot \cos \epsilon n  & -\sin \theta_s n \cdot \sin \epsilon n  & \cos \theta_s n 
 \end{pmatrix}\, . \label{eq:FullEvolution}
 \end{split}
 \end{equation}	 
At the boundary condition $\langle\vec{S}(0)\rangle = P_y^{\rm d}(0) \vec{e}_y$ one has 
 \begin{eqnarray}
 \begin{split}
 \vec{S}(n)  = S_y(0)[&\cos (\epsilon n) \vec{e}_y \\
 + &\sin(\epsilon n)  [ \cos (\theta_s n) \vec{e}_x - \sin (\theta_s n) \vec{e}_z ]\,, \label{eq:Vector}
 \end{split}
 \end{eqnarray}
with the conspicuous interpretation of $\cos \epsilon n$ and $\sin\epsilon n$ being the envelopes of the vertical and in-plane polarizations,  respectively. Apart from the time dependence of the envelope, the in-plane polarization continuously precesses with the idle precession frequency. The in-plane polarization and the idle precession frequency $f_s$ can be determined from the up-down asymmetry, measured in elastic scattering of deuterons on carbon in the polarimeter\,\cite{Bagdasarian:2014ega}. 
The interaction of protons of 
about
 \SI{135}{MeV} 
 with a deuterium target\cite{Valdau:2011zz,Lenisa:2019cgb} amounts to the interaction of deuterons of
 \SI{270}{MeV}
 with  a proton target, incidentally close to the optimum energy for the polarimetry of deuterons,
  elastically scattered off a carbon target\,\cite{2020arXiv200307566J}. 
 
As stated above, the spin-tensor operator does not enter per se the spin-interaction Hamiltonian. For that reason, its evolution can be written down right away, 
 \begin{equation}
 \matr{Q}(n) = \matr{R}_\text{evol} (n) \, \matr{Q}(0) \, \matr{R}_\text{evol}^{\rm T}(n) \,, \label{eq:TensorEvolution}
 \end{equation}
 without the need to solve a set of five differential equations for the tensor polarization operator\,\cite{Huang:1993qj,Baryshevsky:1993hk,Silenko:2015qfa}. Upon averaging over the injected ensemble of vertically polarized particles, one obtains
 \begin{equation}
 P^{\rm d}_{x,z}(0)= \langle S_{x,z} (0) \rangle =0 \, ,\label{eq:VectorZero}
 \end{equation}
 although the in-plane polarizations do build up under the action of the spin rotator [see Eq.\,(\ref{eq:Vector})]. On exactly the same footing, all the off-diagonal tensor polarizations vanish as well, $T_{yx}(0) = T_{yz}(0) =  T_{xz}(0) =0$, and $T_{xx}(0) =  T_{zz} (0)= -\frac{1}{2}T_{yy}(0)$. Then, Eq.\,(\ref{eq:TensorEvolution}) yields
\begin{equation}
\begin{split}
  T_{yy}(n) & = \phantom{-}\frac{1}{2}T_{yy}(0)  \cdot \left[ -1 + 3 \cos^2 \epsilon n \right] \,,\\ 
  T_{xx}(n) & = \phantom{-}\frac{1}{2}T_{yy}(0)  \cdot \left[ -1 + 3 \sin^2 \epsilon n \cdot  \cos^2\theta_s n \right] \,, \\ 
  T_{zz}(n) & = \phantom{-}\frac{1}{2}T_{yy}(0)  \cdot \left[ -1 +3 \sin^2 \epsilon n \cdot \sin^2\theta_s n \right]\,, \\
  T_{yx}(n) & = \phantom{-}\frac{3}{2}T_{yy}(0)  \cdot \sin \epsilon n \cdot \cos \epsilon n \cdot \cos\theta_s n\,, \\ 
  T_{yz}(n) & = -\frac{3}{2}T_{yy}(0) \cdot \sin  \epsilon n \cdot \cos \epsilon n \cdot \sin\theta_s n\,, \\
  T_{xz}(n) & = -\frac{3}{4}T_{yy}(0) \cdot \sin^{2} \epsilon n \cdot \sin 2\theta_s n\,. 
\end{split}
\label{eq:Qyy-Qzz}
\end{equation}
The working point of the experiment, with the vector polarization in the ring plane,  is reached at $n=n^*$, \textit{i.e.}, when $\epsilon n^* = \pi/2$, at which point the RF spin rotator is turned off.  The overall spin-evolution operator upon subsequent $m$ idle precession turns will be  
 \begin{equation}
 \begin{split} 
 \matr{R}_\text{evol}(n^*,m) &= 
 \matr{R}_\text{idle}(m) \, \matr{R}_\text{idle}(n^*) \, \matr{R}_\text{env}(n^*)\\
 & = \matr{R}_\text{idle}(n) \, \matr{R}_\text{env}(n^*)\, ,
 \end{split}
 \end{equation}
 where $n=m+n^*$. This representation allows an easy interpretation: the evolution of envelopes freezes at $n=n^*$, while the idle precession continues without interruption.
  
 Some observations on these results are in order. The tensor polarization evolves from $T_{yy}(0)$ to $T_{yy}(n\geq n^*) = -T_{yy}(0)/2$, and  $T_{xx}(n \geq n^*)+T_{zz}( \geq n^*) = -T_{yy}(n^*)$ does not depend on the spin-precession phase $n\theta_s$. The two off-diagonal tensor polarizations $T_{yx} (n)$ and  $T_{yz}(n)$ can be combined into the in-plane vector, 
 $ \vec{Q}_1(n) = T_{yx}(n)\vec{e}_x +  T_{yz}(n)\vec{e}_z\, ,$
which precesses with the same idle precession frequency as the in-plane vector polarization (see also\,\cite{Huang:1993qj}). Its envelope is the product of the envelopes for $P_y$ and $P_{x,z} $, and it vanishes for pure in-plane vector polarization. Similarly, 
$\vec{Q}_2(n) = \left[ T_{xx}(n) - T_{zz}(n) \right] \vec{e}_x + 2T_{xz}(n)\vec{e}_z$
is the in-plane vector polarization, which precesses with twice the idle precession frequency (see also\,\cite{Huang:1993qj}). Its envelope  $\sin^{2}\epsilon n$ is equal to the square of the envelope of the in-plane vector polarization.
  
The above analysis suggests that the TVPC asymmetry, 
\begin{equation}
A_{\rm TVPC}(n) =-\frac{3}{4} \cdot \frac{\sigma_{\rm TVPC}}{\sigma_0}T_{yy}(0)P_y^{\rm p}\cdot \sin^{2} \epsilon n^* \cdot \sin 2\theta_s n\, ,
\label{eq:TVPCasymmetry}
\end{equation} 
constitutes a unique $P_y^{\rm p}$-odd asymmetry, which oscillates with twice the idle precession frequency.  It is readily distinguishable from the oscillating component of the T-conserving and $P_y^{\rm p}$-independent tensor asymmetry
arising
 from $\sigma_{\rm T} T_{zz}(n) \propto \sin^{2} \epsilon n^*(1- \cos 2 \theta_s n)$. 

In the conventional approach with  static polarizations, the residual $P_{y,\text{res}}^\text{d}$ in the $T_{xz}$ tensor-polarized deuterium target, contributes a hard to quantify $P_y^\text{p}$-odd background $\propto  \sigma_\text{TT} \, P_{y,\text{res}}^\text{d}\, P_y^\text{p}$\,  \cite{Valdau:2011zz,Eversheim:2017zxl,Temerbayev:2015foa}. In the novel approach described here, the departure of $\epsilon n^*$ from $\pi/2$ shall give rise to a constant $P_y^\text{p}$-odd offset $\propto \sigma_\text{TT}  P_y^\text{p} P^{\rm d}_y(0) \cos \epsilon n^*$,  which does not affect the determination of the oscillating TVPC term by a Fourier analysis.
 
A misalignment of the proton polarization in the target with respect to the normal to the ring plane $\vec{e}_y$ generates unwanted  $P_{x}^\text{p}$ and $P_{z}^\text{p}$ components [see also\,\cite{Rathmann:1998zz}]. These will produce a background $\vec{P}^\text{p}$-odd signal 
$\propto P^{\rm d}_y(0) \sin \epsilon n^*\,(\sigma_\text{TT} \, P_{x}^\text{p}\,\cos\theta_s n -\sigma_\text{LL}\,P_z^p\,\sin\theta_s n)$, distinguishable from the TVPC signal by Fourier analysis. A further unwanted contribution from misalignment is the $P_x^\text{p}$-odd term $\sigma_\text{TVPC} T_{yz}(n) \, P_x^\text{p} \propto  \sin  \epsilon n^* \cdot \cos \epsilon n^* \cdot \sin\theta_s n$ [see Eq.\,(\ref{eq:TVPC})]. It is part of the TVPC signal that oscillates with the idle precession frequency $f_s$. Besides that, this signal is suppressed because $\cos\epsilon n^*$ is small.  
 
Yet another background from misalignment stems from the P-odd tensor cross section, $\sigma_{\rm PV}^\text{T} \,  P^\text{p}_z \, T_{zz}(n) \propto  P^\text{p}_z\cdot (1- \cos 2 \theta_s n)\cdot\sin^{2} \epsilon n^*$. Besides  $P^\text{p}_z$ being small, this contribution is additionally suppressed by parity violation.
 
The first approach to isolate $\sigma_{\rm TVPV}({\bf k}\cdot [{\bf P}^{\rm d}\times {\bf P}^{\rm p}])$ demands/ for the radial polarization of target protons. The technique of switching the target polarization  in sign and direction by reversing a weak guide field is described in\,\cite{Rathmann:1998zz}. In this case 
\begin{equation}
\sigma_{\rm TVPV}{\bf k} \cdot \left[ {\bf P}^{\rm d}\times {\bf P}^{\rm p} \right] \propto P_{y}^{\rm d}(0) \, P_x^{\rm p}\cos\epsilon n\, ,
\end{equation}
the corresponding T-odd signal will be $P_x^{\rm p}$-odd and has a characteristic envelope $\cos \epsilon n $. It is easily  distinguishable from the oscillating $\sigma_\text{TT} \, P_x^\text{d}(n)\, P_x^\text{p} \propto \sin \epsilon n^* \cdot \cos \theta_s n$.

The second approach to the TVPV asymmetry is to retain the vertical polarization of protons and to take advantage of the precessing vector polarization of deuterons. In that case, 
\begin{equation}
\begin{split}
{\bf k} \cdot \left[ {\bf P}^\text{d} \times {\bf P}^\text{p} \right] =  P_{x}^{\rm d}(n) \, P_y^{\rm p}  \propto P_y^{\rm p} \,\sin\epsilon n^* \cdot \cos\theta_s n \, 
\label{eq:TVPV_Py}
\end{split}
\end{equation}  
is $P_y^{\rm p}$-odd,  and oscillates with the frequency $f_s$. Furthermore, choosing a second working point, $\epsilon n^* = 3\pi/2$, offers an extra cross check of the T-violation property, as it  amounts to the reversal of the sign of the deuteron in-plane vector polarization compared to the case of $n^* = \pi/2$. Thus, the oscillating signal (see\,\ref{eq:TVPV_Py}) can readily be isolated from the offset $\sigma_\text{TT} \, P_y^\text{d}(n^*) \, P_y^\text{p}$.

The case of the TVPV asymmetry summarizes the remarkable power of the precessing-polarization approach to T-violation in $pd$ interactions. In the same experimental setup, one can \textit{simultaneously} search for TVPC and TVPV asymmetries with about the same sensitivity.
 
Next we comment on longitudinal spin asymmetries. In the case of an unpolarized target, the P-violating asymmetry $A_\text{PV}^\text{d}(n) \propto \sin \theta_s n$ oscillates with the frequency $f_s$, and the method is equally applicable to orbiting horizontally polarized protons\,\cite{2019arXiv191110701K}.  
Apart from $A_\text{PV}^\text{d}(n)$, there will be an extra signal from the tensor polarization $T_{zz}(n)$, which has a component oscillating with the frequency $2f_s$. The two signals are readily separated by the Fourier analysis. This possibility to measure the tensor asymmetry $A_\text{T} = \sigma_\text{T}/\sigma_0$ in the scattering of deuterons on various unpolarized internal targets comes as an extra bonus from the in-plane precessing deuteron polarization. 

To measure the P-conserving double longitudinal asymmetry $A_\text{LL}$ with stored deuterons, one needs longitudinally polarized protons in the target cell \cite{Rathmann:1998zz}. For the deuteron spins in the ring plane, the observed asymmetry equals
\begin{equation}
A_\text{LL}^{\rm dp,pp}(n) =  \frac{\sigma_\text{LL}}{\sigma_0} \, P_z^\text{d}(n) \,  P_z^\text{p} \propto  P_z^\text{p} \, \sin \theta_s n\, . 
\end{equation} 
It is $P_z^\text{p}$-odd and oscillates with the frequency $f_s$. The same technique applies to stored polarized protons.
 
Alongside $ A_\text{LL}(n)$, one will gain access to the PV tensor asymmetry $A_\text{PV}^\text{T}$. It emerges from the dumbbell structure of the deuteron as soon as parity is violated in $pp$ and $pn$ scattering. The signal of this asymmetry will be $P_z^\text{p}$-odd and will oscillate as $\cos \theta_s n$. A conservative expectation for this asymmetry is $A_\text{PV}^\text{T} \approx A_\text{PV}^\text{p} \, A_\text{T} \sim 10^{-7}\times A_{\rm T}$. To the best of our knowledge, this asymmetry has never been searched for experimentally, and the issue of $A_\text{PV}^\text{T}$  is an entirely open one.

 We presented a new approach to search for T-violation in $pd$ interactions in a storage ring experiments. The Fourier analysis of oscillating spin asymmetries, in conjunction with the reversal of the proton target polarization, will enable one to uniquely determine the T-violating and P-conserving asymmetry $A_\text{TVPC}$, and simultaneously the T-violating and P-violating asymmetry $A_\text{TVPV}$. The desired oscillating tensor and vector polarizations can be generated by the RF driven vertical-to-in-plane rotation of the deuteron polarization in a storage ring. Based on the developed simple analytic description of the evolution of the tensor polarization, we have shown that, as a byproduct, the same technique of precessing vector polarization can be used to study concurrently in the same experimental setup a whole family of P-odd and P-even spin asymmetries.  
 
 N.N.N.\ was supported by the Russian Fund for Basic Research (Grant No. 18-02-40092 MEGA). A.S.\ acknowledges his support by the NNSF of China (Grants No.\ 11975320 and No.\ 11805242), and by the CAS President’s International Fellowship Initiative (No.\ 2019VMA0019), and the hospitality at the Institute of Modern Physics of the CAS. F.R.\ has been supported in the framework of an ERC Advanced-Grant of the European Union (proposal number 694340).
 
\bibliographystyle{apsrev4-1}
\bibliography{spintunemapping_23.01.2017}

\end{document}